%% file: main.tex
\documentclass[
twocolumn,
]{ceurart}

\usepackage{listings}
\usepackage{import}
\usepackage{CJKutf8}
\usepackage{rotating}
\usepackage{multirow}
\usepackage{makecell}

\begin{document}

\copyrightyear{2023}
\copyrightclause{Copyright for this paper by its authors. Use permitted under Creative Commons License Attribution 4.0 International (CC BY 4.0).}
\conference{Fifth Knowledge-aware and Conversational Recommender Systems (KaRS) Workshop @ RecSys 2023, September 18--22 2023, Singapore.}


\title{MeKB-Rec: Personal Knowledge Graph Learning for Cross-Domain Recommendation}


\author[1]{Xin Su}[%
orcid=0000-0002-5979-8702,
email=levisu@tencent.com
]
\fnmark[1]

\author[1]{Yao Zhou}[%
email=yaoozhou@tencent.com
]
\fnmark[1]

\author[1]{Zifei Shan}[%
orcid=0000-0002-8283-6498,
email=zifeishan@tencent.com
]
\cormark[1]
\fnmark[1]

\author[1]{Qian Chen}[
email=qiancheng@tencent.com
]

\address[1]{WeiXin Group of Tencent, ShenZhen, China}

\cortext[1]{Corresponding author.}
\fntext[1]{These authors contributed equally.}


\begin{abstract}
It is a long-standing challenge in modern recommender systems to make recommendations for new users, namely the cold-start problem. Cross-Domain Recommendation (CDR) has been proposed to address this challenge, but current ways to represent users' interests across systems are still severely limited.
We introduce Personal Knowledge Graph (PKG) as a domain-invariant interest representation,
and propose a novel CDR paradigm named MeKB-Rec. 
We first link users and entities in a knowledge base
to construct a PKG of users' interests, named MeKB.
Then we learn a semantic representation of MeKB for the cross-domain recommendation.
Beyond most existing systems, our approach builds a semantic mapping across domains using Pretrained Language Models which breaks the requirement for in-domain user behaviors, enabling zero-shot recommendations for new users in a low-resource domain.
We experiment MeKB-Rec on well-established public CDR datasets,
and demonstrate that the new formulation achieves a new state-of-the-art that significantly improves HR@10 and NDCG@10 metrics over best previous approaches by 24\%--91\%, with a 105\% improvement for HR@10 of zero-shot users with no behavior in the target domain. 
We deploy MeKB-Rec in WeiXin recommendation scenarios and achieve significant gains in core online metrics. MeKB-Rec is now serving hundreds of millions of users in real-world products.
\end{abstract}


\maketitle

\section{Introduction}
\label{sec:intro}
\import{./}{2-introduction.tex}

\section{Related Work}
\import{./}{3-related.tex}

\section{Approach}
\import{./}{4-approach.tex}

\section{Experiments}
\import{./}{5-experiments.tex}

\section{Conclusion}
\import{./}{6-conclusion.tex}
\bibliography{reference}


\appendix

\section{Dataset Details}
\import{./}{10-appendix1.tex}

\section{Additional Experiments}
\label{appendix:exp}

\import{./}{11-appendix2.tex}



\end{document}

%% file: 2-introduction.tex
The cold-start problem has been a core challenge for online Recommender Systems for decades.
One promising line of methods to deal with the cold-start problem is Cross-Domain Recommendation (CDR) \cite{schein2002methods, zang2021survey} , which transfers user and item information from other domains to assist recommendation for the target domain \cite{li2020ddtcdr,liu2020cross, hu2018conet}.
In this paper, we focus on a common scenario in CDR, where users partially overlap in two domains, and many users have rich behaviors in the source domain but no or few interactions in the target domain. 
How to effectively make recommendations for such users is a key challenge to improve modern Recommender Systems.

\begin{figure}[t]
  \centering
  \includegraphics[width=0.95\linewidth]{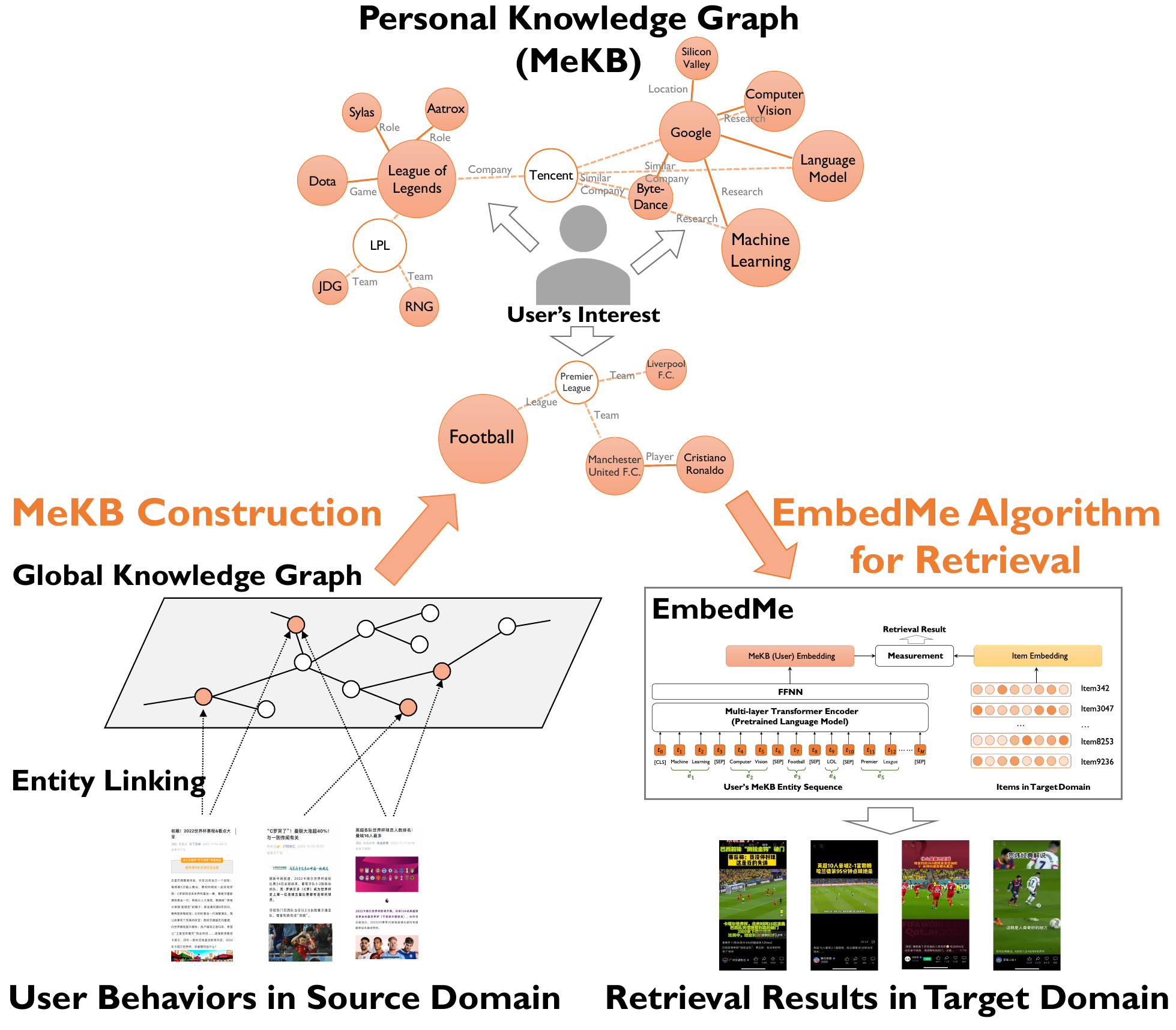} 
  \caption{The MeKB-Rec system consists of two steps. The MeKB construction step builds a personal knowledge graph (MeKB) representing a user's interest as a sub-graph of entities. The EmbedMe algorithm step employs a pre-trained language model to map the entity sequence of MeKB for retrieval in the target domain. }
  \label{Figure1}
  \vspace{-0.45cm}
\end{figure}

The central problem for CDR lies in learning transferable representations with the help of users that overlap between domains.
However, most existing CDR approaches are constrained by a wide-spread hypothesis that 
embedding representations of users can be sufficiently learned from a few examples of overlapped users.
We challenge this hypothesis by raising that, 
1) the user representation with merely an embedding has limited transferability,
as it tends to weaken the diverse semantics behind users' interactions in the source domain;
2) the data sparsity for overlapped users can severely hinder the learning effectiveness, without help of external knowledge.


To this end, we introduce the concept of Personal Knowledge Graphs (PKGs), originally proposed by \citet{yang2022pkg} to model relations between users and real-world entities.
We propose an interest-oriented PKG that models interest relations between users and entities, named \textbf{MeKB}.
At the core of its advantages,
(1) MeKB is semantically diverse, in that it leverages external information from large Knowledge Graphs (KGs) 
to connect and organize discrete user interests in the source domain;
(2) MeKB is transferable, in that it forms a domain-agnostic graph of user interests that are semantically stable for cross-domain transfers.



To build MeKB, we first perform Entity Linking (EL) \cite{ling2015design, zhang2021entqa} on items' textual content to identify their entities of interest. 
Then we connect the entities with users through user-item interactions  (e.g. likes or reviews) in the source domain.
These steps establish MeKB, a sub-graph for the global KG centered around the user to express their interests.
MeKB is essentially a general interest storage that represents user interests as world entities.


With MeKB constructed, it remains non-trivial to make cross-domain recommendations with this rich user representation.
We notice that Pretrained Language Models (PLMs) are recognized as good few-shot learners \cite{brown2020language, gao-etal-2021-making}.
Here we propose \textbf{EmbedMe}, a representation learning mechanism to map semantic knowledge in MeKB to fit recommendation tasks in the target domain. 
By combining MeKB with PLMs, EmbedMe can elevate the semantic knowledge to significantly ease the learning process in CDR, a novel approach to bring world knowledge in PLMs into recommender systems.
In particular,
EmbedMe first flattens MeKB into a weighted sequence of entities, 
and then encodes the sequence with a PLM-powered transformer encoder to retrieve recommendation results for the target domain.
The construction of MeKB and the EmbedMe mapping model together form the MeKB-Rec system, as illustrated in \autoref{Figure1}.


We conduct experiments with MeKB-Rec on two real-world recommendation products.
We first benchmark on three settings in the Amazon dataset which is commonly used for CDR evaluations \cite{ni2019justifying, cao2022disencdr}.
Compared with prior methods, MeKB-Rec achieves the state-of-the-art performance, outperforming other systems by a large margin on HR@10 and NDCG@10 metrics (varying from \textbf{24\%} to \textbf{91\%}).
We will publish this processed dataset to facilitate future research in the camera-ready.
Furthermore, we conduct offline and online experiments on WeiXin, a popular online platform with billions of users. 
Besides improvements in offline metrics,
MeKB-Rec provides significant DAU and PCTR gains in online A/B testing of WeiXin Live recommendation. 
MeKB-Rec is now deployed into the product to serve hundreds of millions of users.

%% file: 3-related.tex
Recommendation systems have widely become the key component for many online information systems in the past decades. Two key challenges for the growth of recommendation systems are, 
(1) how to effectively recommend where user-item interactions are limited, namely the data sparsity problem; and
(2) how to handle new users or items that enter the system, namely the cold-start problem. \cite{zhu2019addressing}
To address these problems, Cross-Domain Recommendation (CDR) is an increasingly popular direction.
The setting of CDR involves two recommendation domains, namely ``source'' and ``target'', and the key idea is to transfer users' representations from the source domain to assist recommendation in the target domain.
\cite{da2020recommendation}

According to different recommendation settings, CDR approaches are divided into sub-problems: user (or item) non-overlap, user (or item) partial overlap and user (or item) full overlap \cite{zang2021survey}.
This paper studies the scenario of user partial overlap, which is an important case that occurs frequently in real-world recommender systems.

In the user partial-overlapped scenario, existing CDR approaches typically focus on building feature representations of users that can carry over from the source to the target domain.
One common approach is Embedding \& Mapping.
Specifically, a user's feature in the source domain is represented by an embedding $U_i^s$, and the user feature in the target domain is represented as $U_i^t$.
During the transfer learning process, the model is trained by overlapped users to learn a mapping function $f_U$, which maps the feature $U_i^s$ into a downstream-friendly representation in the target domain. \cite{man2017cross}.
The object is to establish the relationship between the latent space of domains:
\begin{equation}
\min_\theta \sum_{u_i \in \mathbb{U}} L(f_U(U_i^s; \theta), U_i^t).
\end{equation}
Embedding-and-mapping approaches aim at transferring representations across domains \cite{zhang2020learning, li2021dual, zhu2021transfer}, but the single feature embedding become the bottleneck for expressing a user's diverse interest. Besides, learning the embedding requires a large amount of examples for overlapped users.
In real-world CDR problems, the variety of user features and the lack of overlapped users' interactions can limit the application scope of the Embedding \& Mapping method.

To utilize more user features across domains beyond the traditional feature embedding, another approach is building shared graphs to represent the relationships among users, items, attributes, and other factors.
The constructed graph integrates the information over both source and target domains, and by using Graph Neural Networks, the users' representations are embedded into a shared latent space \cite{wang2017item, cui2020herograph, li2020heterogeneous, xu2021expanding}.
Although such methods successfully introduce more features in the source domain and improve CDR performance, they tend to ignore the rich semantics behind users' discrete interest points, and the insufficient amount of overlapped users' behavior leads to significant difficulty for optimization.

The goal of this paper is to break the limitation of traditional user representations
in existing CDR approaches.


%% file: 4-approach.tex
In this section, we demonstrate the proposed approach MeKB-Rec, a two-stage recommendation system, as illustrated in \autoref{Figure1}.
The first stage is building MeKB, a personal knowledge graph that connects users and entities via interest relations, derived from interactions in the source domain.
The second stage is EmbedMe, an effective
representation learning model that maps MeKB to a downstream embedding for retrieval 
in the target domain.

\subsection{MeKB Construction}

\begin{figure}[ht]
  \centering
  \includegraphics[width=0.8\linewidth]{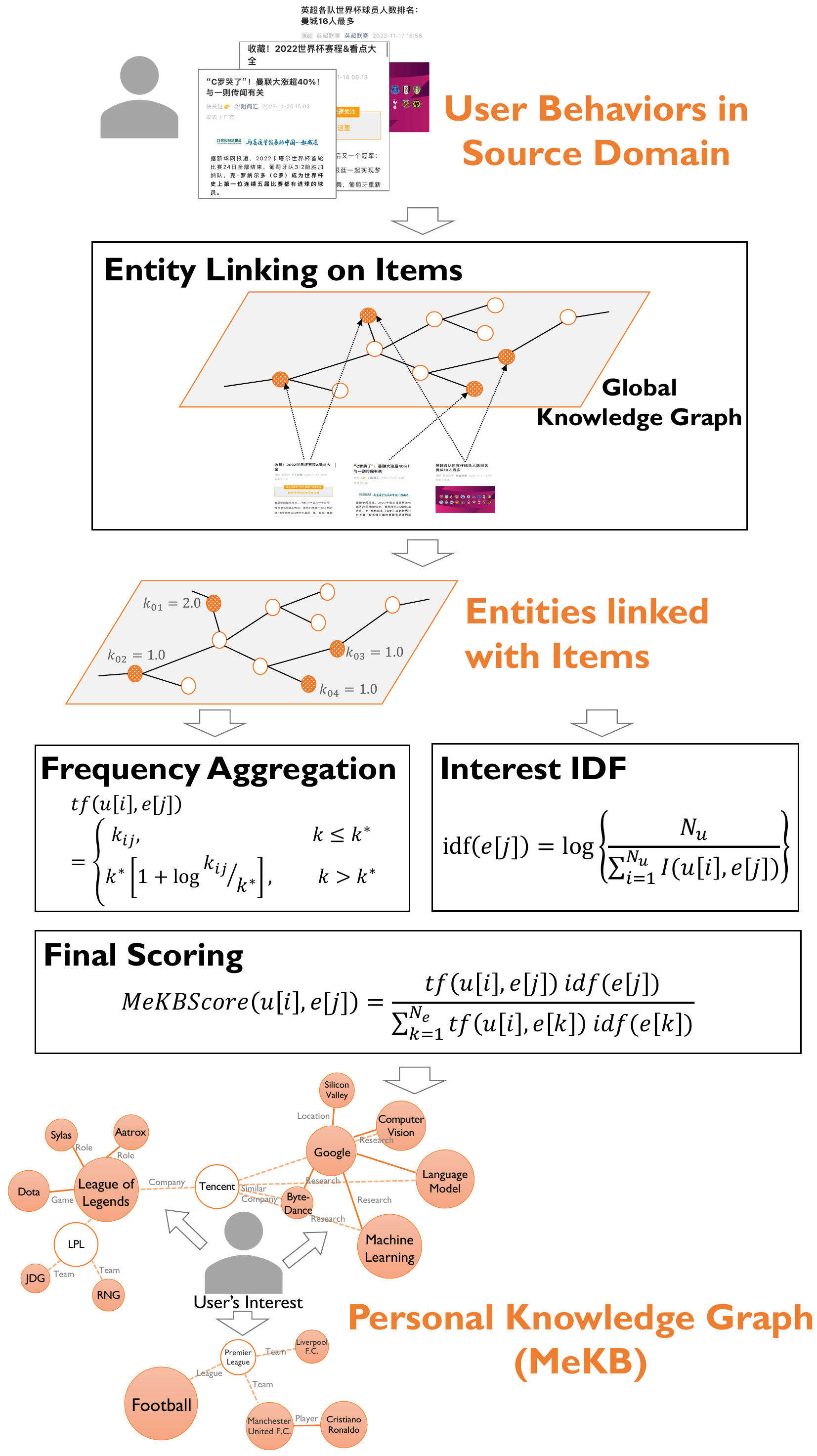}
  \caption{MeKB is constructed based on users' behaviors in the source domain. We first extract a subgraph of the global KG by performing entity linking on items. Linked entities are weighted by Frequency Aggregation on user-item interactions, and an Interest IDF scoring function. 
  }
  \label{Figure2}
  \vspace{-0.4cm}
\end{figure}

MeKB is an instance of Personal Knowledge Graphs (PKGs) \cite{balog2019personal}. It distills user-item interactions into user-entity interest relations by mapping items into entities of interest.
Different with prior works that learns latent interest representations, MeKB is an explicit interest store for a user's interests in the form of entities.
The overall process for MeKB construction is shown in \autoref{Figure2}.

\subsubsection{Knowledge Graph.}

Knowledge Graphs (KGs) or Knowledge Bases (KBs) were proposed to represent structural world knowledge \cite{vrandevcic2014wikidata, chen2020review, zhang2014feature}, and have been widely adopted for applications such as Search and Question Answering \cite{fevry2020entities}. There have been efforts to use KGs for recommendation \cite{wang2019explainable, wang2021learning}, but few work used KGs in the CDR setting.
Personal Knowledge Graphs (PKGs) were proposed in 2019 \cite{balog2019personal} as a new concept to model user-entity interactions. Some work adopted PKGs for recommendations \cite{yang2022pkg}, though most work focused on news recommendations and few on the CDR problem.

In this paper, we use Wikidata \cite{vrandevcic2012wikidata} as our KG.
The set of all entities in KG is denoted as $E$. Each $e \in E$ is an entity: persons, organizations, locations, works etc, usually with a Wikipedia page.
The set of relations between entities is denoted as $R$. Each $r \in R$ is a triple between two entities: $(e1, rtype, e2)$ where $e1 \in E$ and $e2 \in E$.
Each entity comes with a unique title $Title(e)$ for the given language.
Note that in principle, one can replace Wikidata with another KG or augment it with domain-specific KGs to build MeKB.



\subsubsection{Entity Linking on Items.}

Built upon Wikidata as the global KG, the next step is to connect items with entities of interest in KG.
Entity Linking (EL) is the task that links texts with entities in KG. In MeKB construction, it serves as the key step to connect recommendation items with KG to build its semantic representation.

We formalize the definition of Entity Linking in our scenario. Given KG entities $E$ and a recommendation item $i \in I$, denote the textual content of the item as $Text_i$. An Entity Linking system outputs as set of entities: 
$\{e_1, e_2, \dots, e_n\}$, where $\forall_{j \in [1,n] } e_j \in E$. Output entities are correlated with the item text, denoted $HasMention$:
\label{eq:hasmention}
\begin{equation}
\forall_{j \in [1,n]} HasMention(Text_i, e_j)
\end{equation}


In this paper, we use off-the-shelf EL methods for MeKB construction.
We follow previous work \cite{genre, fevry2020empirical, xu2022hansel} to extract an Alias Table (AT) from Wikipedia, by parsing Wikipedia internal links, redirections, and page titles. 
An Alias Table defines the prior probability of a text mention $m$ linking to an entity $e$, estimated with the formulation:
\begin{equation}\label{eq:prior}
    P(e|m) = \frac{freq(m,e)}{freq(m)},
\end{equation}
where $freq(m,e)$ denotes the number of anchor texts with the surface form $m$ with reference to the entity page $e$ in Wikipedia, and $freq(m)$ denotes the total number of anchor texts with the surface form $m$.

To enable application on a large quantity of items, for the sake of simplicity and efficiency, we directly use the top-1 result from the Alias Table (AT@1) for disambiguation, which is considered a good EL benchmark across languages \cite{el100}. 

As our experiment scenarios span English and Chinese languages, we choose available EL methods in both languages:
for English datasets, we first perform named entity recognition (NER) on the textual content of the item, powered by Spacy \cite{spacy}, then perform AT@1 matching \cite{genre, xu2022hansel} with the constructed alias table to map the name to the entity. 
For Chinese datasets, we directly match against the item's textual content against the alias table as released in \citet{xu2022hansel}.



\subsubsection{User-Item Interaction Aggregation.}

Naively, by aggregating all positive user-item interactions and collecting entities on items, we can already build the first version of MeKB.
However, naive frequency-based aggregations assume that entities appearing more frequently better represent users' interest, which is often inaccurate in actual recommender systems.
To make MeKB better match the characteristics of real-world scenes,
we devise an aggregation mechanism to adjust entity weights in the personal knowledge graph.

Concretely, the global KG reflects the global knowledge of the world. KG entities have equal rights, and the edges between entities only reflect the objective world relations.
However, for the recommendation system, it remains challenging to model which entities play a better role in representing users' interests. 
On one hand, counting the number of items that associate with each entity gives a first estimation of the interest's strength. 
On the other hand, this frequency-based measurement may not be ideal as it can overestimate entities with higher global frequency.
Therefore, for these two factors, we respectively integrate two interaction aggregation methods into the entity weight of MeKB: 1) Frequency Aggregation and 2) Interest IDF Scoring. 


\textbf{1) Frequency Aggregation.} We first set a frequency score for each entity in the original personal knowledge graph, according to the number of user-item interactions that are associated with each entity. 
For example, a user clicks on $k_1$ videos associated with entity \textit{Association\_football} and $k_2$ videos with \textit{Manchester\_United\_F.C.}.
Then the initial frequency scores of the corresponding entities are set to $k_1$ and $k_2$.
Considering that some entities with extra-high frequency may dominate the head distribution, we perform a log smoothing for the frequency greater than a certain threshold K, that is
\begin{equation}
s(k) = 
\begin{cases}
k & k \leq k\text{*} \\
k \big [1 + \log{\frac{k}{k*}} \big ] & k>k\text{*}
\end{cases}
\end{equation}

\textbf{2) Interest IDF Scoring.} 
Considering the deviation of popularity distribution in actual recommendation scenarios,
it is beneficial to reduce the weight of entities with higher exposure so that the scores in MeKB are more in line with the actual strength of users' interests.
We are inspired by the concept of IDF \cite{qaiser2018text} from natural language processing, to judge the contribution of an interest entity in recommendation.  
Intuitively, IDF is correlated with information gain. An IDF-weighted interest score downweights popular interests and reduces exposure bias on the recommendation system. 
The IDF calculation formula for each entity is as follows.
\begin{equation}
idf(e[j]) = \log \Big \{ N_u / \sum_{i=1}^{N_u} I(u[i],e[j]) \Big \}
\end{equation}
where $u[i]$ is the user $i$ and the $e[j]$ is the entity $j$, and $I$ is an indicator function for whether the interest relationship between $u[i]$ and $e[j]$ exists. 
We note that $tf(u[i],e[j])$ is the cooccurrence frequency of $u[i]$ and $e[j]$ as calculated by the Frequency Aggregation step, that is
\begin{equation}
tf(u[i],e[j]) = s(k_{ij}) 
\end{equation}
Then we perform IDF on the entity frequency score to obtain the final interest score of MeKB, denoted $MeKBScore$:
\begin{equation}
MeKBScore(u[i],e[j]) = \frac{tf(u[i],e[j]) \cdot idf(e[j])} { \sum_{k=1}^{m} tf(u[i],e[k]) \cdot idf(e[k]) }
\end{equation}

In short, the Interest IDF and Frequency Aggregation scores are combined with multiplication to compute the final interest weight for each entity in MeKB.

\begin{figure*}[ht]
  \centering
  \includegraphics[width=1.0\linewidth]{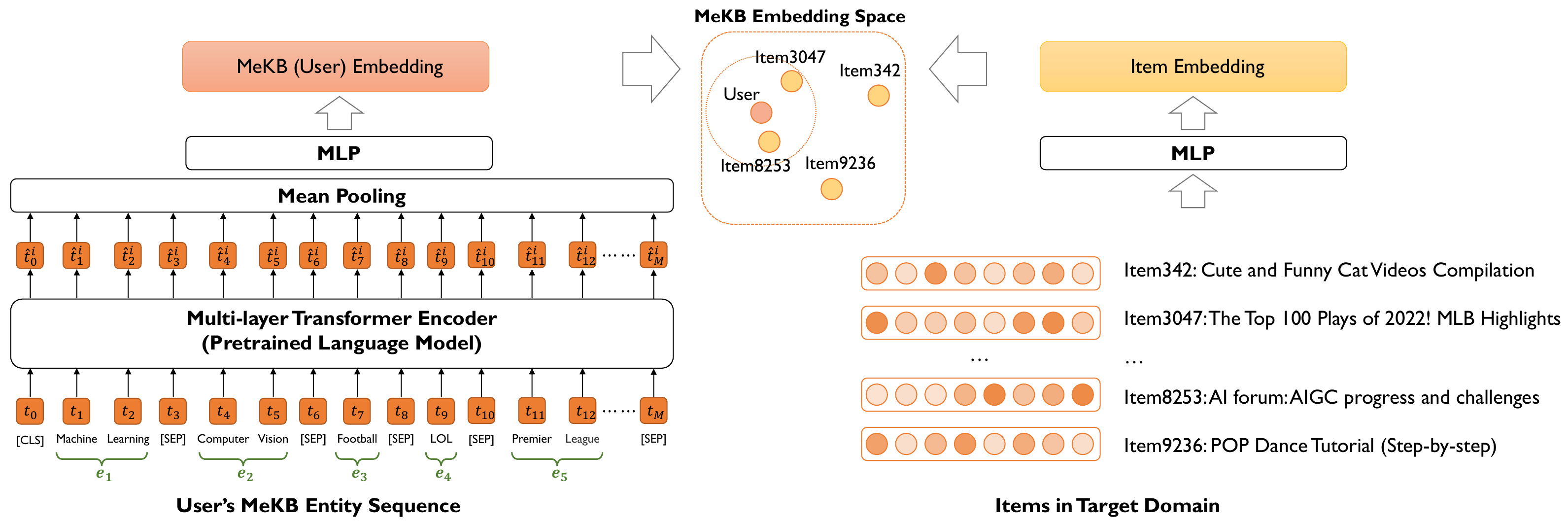}
  \caption{EmbedMe model structure. The user encoder is a PLM on the transformed MeKB sequence input, followed by an MLP projecting to a user embedding of dimension $K$. The item is simply represented by its ID embedding. Two towers are jointly optimized for retrieval over positive examples in the target domain.}
  \label{Figure3}
  \vspace{-0.5cm}
\end{figure*}

\subsection{EmbedMe Algorithm}
Although MeKB makes user interest semantically valid across domains, there is still an essential difficulty for learning to make accurate recommendations from MeKB to items in the target domain.
The critical challenge is that, although MeKB adopts large knowledge graphs to bring in a large amount of auxiliary entity information and interest associations, the scarcity of CDR samples makes it difficult to implement end-to-end training with this amount of auxiliary features.
On one hand, MeKB has a holistic view of a user's interests as a set of entities in the global KG, each of which is semantically meaningful.
On the other hand, the overlapped users' behavior data is not enough to support end-to-end training of recommendation systems for all PKGs in the CDR setting.

To overcome this challenge, we present an algorithm to map a user's personal knowledge graph, MeKB, into a dense embedding representation suitable for retrieval in the target domain. We name the algorithm EmbedMe.

To implement EmbedMe, we transform the personal knowledge graph into a semantically weighted sequence representation of entities, and then introduce a Pretrained Language Model (PLM) as the semantic encoder to map the entity sequence to an embedding. 
We fine-tune this semantic encoder together with the item representation in the target domain via a dual-encoder architecture.


\subsubsection{Pretrained Language Model}


Natural Language Processing has been enormously successful thanks to the help of Pretrained Language Models (PLMs) such as BERT and GPT \cite{devlin2018bert, floridi2020gpt}, 
leading to significant improvements in Search, Advertisement and Assistant \cite{devlin-etal-2019-bert}.
PLMs are known for their few-shot learning ability \cite{gao-etal-2021-making}, which makes them potentially suitable for CDR tasks with limited training examples.


Our core idea is applying PLMs on a textual representation of MeKB to model user interest. 
The advantage of the PLM is that it can absorb a wide range of corpus for robust semantic understanding,
but there is still gap between the pretrained semantic representation and the specific recommendation system.
Therefore, we devise an end-to-end fine-tuning method of the PLM for the CDR-specific scenario to extract task-oriented user embedding from MeKB.

\subsubsection{MeKB Sequence Encoder.}

We employ the PLM and build a user encoder that transforms MeKB to an output embedding of the user, named \textit{MeKB Embedding}.
Since MeKB is inherently a graph structure, it becomes challenging for PLMs to exploit the structure. 
Therefore, we turn a user $u$'s interest graph into a sequence of natural language tokens, ordered by their adjusted score $MeKBScore(u, e)$ in descending order.
By encoding over the sequential input, the PLM can learn to encode cross-interest interactions to better fit recommendation tasks.
Each entity is represented by its unambiguous title $Title(e)$ (Wikipedia title or name in domain-specific KGs), and tokenized via WordPiece following \citet{devlin-etal-2019-bert}. Entities are separated by a special $[SEP]$ token. For efficiency, the entity sequence is truncated to a maximum of $N$ tokens ($N=256$).
We add a mean-pooling layer over PLM's sequence output on all tokens, followed by a Multi-layer Perceptron (MLP). The MLP transforms the output to an embedding dimension of $K$ ($K=256$), which is the user's MeKB Embedding.

\subsubsection{End-to-end Training.}

For end-to-end training of MeKB Embedding with the specific CDR task, 
we employ a dual-encoder framework that aligns MeKB Embedding and the learned item's embedding in the target domain, into the same representation space. The overall training architecture is illustrated in \autoref{Figure3}.

\textbf{Item representation.} Since item representation is not the focus of the work, we choose to represent items simply by its ID embedding, which is modeled through a weight matrix with dimension $[I, K]$, where $I$ is the number of items in the target domain, and $K$ is the same embedding dimension.



\textbf{Measurement.} The user and item embedding are optimized with dot-product similarity. We define the score between a user and item $s$ as their embeddings' dot-product similarity:
\begin{equation}\label{eq:cosine}
    s(u,i) = \phi(u)^T \psi(i),
\end{equation}
where $\phi$ is the MeKB sequence encoder, and $\psi$ is the item encoder, projecting user and item representations into $K$-dimensional vectors respectively.

\textbf{Optimization.} The dual encoder is trained on positive interactions in the target domain, i.e. records in the target domain that have a non-empty MeKB input. In our experiments, MeKB can be constructed with user interactions in both source and target domains, which yields more available examples than traditional settings where only users with at least a number of actions in both domains are selected for training. \cite{cao2022disencdr}

For each positive example, we optimize for full softmax loss \cite{relic, fevry2020entities} over all items with respect to the measurement:
\begin{equation}\label{eq:loss}
    l(u,i) = -log \frac{exp(s(u,i))}{ \sum_{j=0}^{I} exp(s(u,i_j))}
\end{equation}
where $I$ is the number of items in the target domain.

%% file: 5-experiments.tex
In this section, we conduct various experiments to compare the recommendation performance of the proposed MeKB-Rec methodology and other CDR baselines. 
We present the detailed experiment settings, results, and analyses.


\subsection{Dataset}

We compare MeKB with other baseline methods on two cross-domain recommendation datasets, Amazon \cite{ni2019justifying} and WeiXin.

\textbf{Amazon Datasets.} We selected three popular domains in Amazon: (1) Books, (2) Movies and TV, and (3) CDs and Vinyl, abbreviated as \textbf{Books, Movies and Music}. 
We build three CDR scenarios: \textit{Amazon Books $\rightarrow$ Movies}, \textit{Amazon Books $\rightarrow$ Music}, and \textit{Amazon Movies $\rightarrow$ Music}. 
We randomly select 20\% of the overlapped users and place all their records into the test set, to verify our effectiveness on cold-start users. 

\textbf{WeiXin Datasets.} We build large-scale CDR datasets extracted from WeiXin recommendation scenarios.
The dataset comes from real user-item interaction on the WeiXin platform, involving three recommendation domains: Articles, Channels (short videos), and Live (live broadcasts).
We set up two cross-domain recommendation scenarios that are \textit{WeiXin Articles $\rightarrow$ Videos} and \textit{WeiXin Videos $\rightarrow$ Live},
and construct CDR datasets with millions of records. 
The detailed parameters of the Amazon and WeiXin datasets used in the experiments are shown in \autoref{Tab-statis-01}. 
The Overlap Ratio is the fraction of overlapped users of two domains over the number of users in the current domain, i.e. $OR(X, Y) = |U_{X} \cap U_{Y}| / |U_{X}|$.

\begin{table*}[t]
\centering
\caption{Experiments on Amazon CDR Dataset}
\label{Tab-amazon-cdr}
\begin{tabular}{lcccccc}
\toprule
\multirow{2}{*}{Methods} & \multicolumn{2}{c}{Books $\rightarrow$ Movies }                 & \multicolumn{2}{c}{Books $\rightarrow$ Music} &   \multicolumn{2}{c}{Movies $\rightarrow$ Music}              \\
\cmidrule(r){2-3} \cmidrule(r){4-5}  \cmidrule(r){6-7}           & HR @ 10  & NDCG @ 10 & HR @ 10 & NDCG @ 10 & HR @ 10 & NDCG @ 10 \\
\midrule
YouTubeDNN & 0.1271 & 0.0762 & 0.1655 & 0.1134 & 0.1707 & 0.1185 \\
CMF               & 0.1726 & 0.0995  & 0.2510   & 0.1659    & 0.2615    & 0.1529  \\
CoNet & 0.1364 & 0.0835 & 0.1948 & 0.1363 & 0.1758 & 0.1246 \\
EMCDR             & 0.1755 & \underline{0.1103}  & 0.2815 & \underline{0.1819}    & 0.2683    & 0.1605  \\
DCDCSR & 0.1450 & 0.0826 & 0.1981 & 0.1563 & 0.1850 & 0.1479 \\
DDTCDR             & 0.1717 & 0.0805  & 0.2839   & 0.1591    & 0.2956    & 0.1666   \\
SSCDR  & \underline{0.1796} & {0.0919}  & \underline{0.2925}   & {0.1713}    & \underline{0.3070}    & \underline{0.1807}   \\
\midrule
\textbf{MeKB-Rec} & \textbf{0.2471} & \textbf{0.1370}  & \textbf{0.4884}   & \textbf{0.3298}    & \textbf{0.5035}    & \textbf{0.3444}    \\
Improved  & \textbf{+37.58\%} & \textbf{+24.21\%} & \textbf{+66.97\%} & \textbf{+81.31\%} & \textbf{+64.01\%} & \textbf{+90.59\%} \\
\bottomrule
\end{tabular}  
\end{table*}

\begin{table*}[t]
\centering
\caption{Experiments on WeiXin CDR Dataset}
\label{Tab-wechat-cdr}
\begin{tabular}{lcccc}
\toprule
\multirow{2}{*}{Methods} & \multicolumn{2}{c}{
\begin{tabular}[c]{@{}c@{}} WeiXin\\
    Articles $\rightarrow$ Videos 
\end{tabular}  }  & \multicolumn{2}{c}{
\begin{tabular}[c]{@{}c@{}} WeiXin\\
    Videos $\rightarrow$ Live
\end{tabular}}  \\
\cmidrule(r){2-3} \cmidrule(r){4-5}     & 
\begin{tabular}[c]{@{}c@{}}  
    HR@10 \end{tabular} &
\begin{tabular}[c]{@{}c@{}}  
    NDCG@10 \end{tabular}  &
\begin{tabular}[c]{@{}c@{}}  
    HR@10 \end{tabular} &
\begin{tabular}[c]{@{}c@{}}  
    NDCG@10 \end{tabular}\\
\midrule
YoutubeDNN    & 0.0733   & 0.0415     & 0.1375 & 0.0816   \\
\textbf{MeKB-Rec}   & \textbf{0.1568}   & \textbf{0.0842}    & \textbf{0.2919}    & \textbf{0.1855}  \\
\midrule
Improved          & \textbf{+113.92\%}   & \textbf{+102.89\%}    & \textbf{+112.29\%}  & \textbf{+127.33\%}   \\
\bottomrule
\end{tabular}  
\vspace{-0.25cm}
\end{table*}

\subsection{Experiment Settings}

In the experiments, we randomly divide each dataset into training, validation and test sets according to the ratio of 8:1:1. We also put 20\% overlapped users in test set of target domain for cold-start evaluation. In every experiment, we repeat the training and testing process five times and take the average as the final result.
In terms of evaluation methods, we adopt \textbf{Hit Rate @K} and \textbf{NDCG @K} as the evaluation criteria for the recommendation systems. We choose K=10 and sample 1 positive item and 999 negative items to compute the metrics for all datasets.

For the implementation of MeKB-Rec, we distribute a total of 1024 examples into 8 NVIDIA A100 GPUs to accelerate training. We adopt Lamb optimizer \cite{you2019large-lamb} with a weight decay of 5e-4, and set its initial learning rate to 5e-4. We train MeKB-Rec for 15 epochs and choose cosine learning rate schedule with first epoch for warm-up. We set the length of sequence to 256 for all experiments.


\subsection{Baseline}
We compare our proposed MeKB-Rec method with the following state-of-the-art recommendation methods, among which the first two are general recommendation methods, while others are specifically for CDR:

(1) YoutubeDNN \cite{youtubednn} is a popular and efficient DSSM model for recommendation retrieval tasks, not specifically designed for CDR.

(2) CMF: Collective Matrix Factorization \cite{singh2008relational-cmf} adopts a Matrix Factorization (MF) method to simultaneously factorize the rating matrices across the source and target domains, where user representations are shared between two domains.

(3) CoNet: Collaborative Cross Networks \cite{hu2018conet} replaces matrix factorization with neural networks in CDR problem, and builds a cross-mapping to connect hidden layers in two base networks.


(4) EMCDR: Embedding-and-Mapping framework for CDR \cite{man2017cross} is proposed for inter-domain recommendation. The method applies both MF and Bayesian Personalized Ranking (BPR) for generating latent factors for users and items. A mapping function with linear and non-linear MLP is used to project representations into the same space for overlapping users in both domains.

(5) DCDCSR: a Deep framework for both Cross-Domain and Cross-System Recommendations \cite{zhu2020deep-dcdcsr} is proposed to resolve the two challenges. The method is based on Matrix MF models and a fully connected Deep Neural Network (DNN).


(6) DDTCDR: Deep Dual Transfer Cross Domain Recommendation \cite{li2020ddtcdr} uses a pretrained autoencoder for user and item representation.
Users' within-domain preferences and cross-domain preferences are jointly modeled. 
Note that the auto-encoders are pre-trained on recommendation data without incorporating world semantic knowledge.

(7) SSCDR: Semi-Supervised framework for Cross-Domain Recommendation \cite{kang2019semi-sscdr} is proposed to utilize both overlapping users and source-domain items to train the overlapping function.



\subsection{Results}

The evaluation results on Amazon datasets are shown in \autoref{Tab-amazon-cdr}.
Results on WeiXin datasets are in \autoref{Tab-wechat-cdr}.
On all datasets, MeKB-Rec significantly outperforms all other methods by a large margin.

\subsubsection{Amazon Dataset}

On Amazon Datasets, MeKB-Rec achieves uniform improvements on all settings, a 24\%--91\% gain over the second-best methods.
The advantage in the \textit{Books $\rightarrow$ Movies} scenario is smaller than that in the other two scenarios. 
We hypothesize that CDR settings with a larger domain gap or a lower user overlap ratio may observe more advantages by using our method.

We further conduct additional ablation studies to validate the effectiveness on PLMs, and show that the pretrained parameters play a key role on the effectiveness of MeKB-Rec. We conduct a frequency-binned analysis on user activity, and find that MeKB-Rec is particularly effective for zero-shot and few-shot users. These analyses are in \autoref{appendix:exp}.

\subsubsection{WeiXin Dataset}


On the large-scale WeiXin dataset, we compare results between MeKB-Rec and the established YoutubeDNN baseline. \cite{youtubednn} Other methods are not compared due to performance reasons at the scale of this dataset.
As shown in \autoref{Tab-wechat-cdr}, on both CDR settings tested, MeKB-Rec performs significantly better than YoutubeDNN,
and the margin of advantage is larger than that of the Amazon dataset.

\subsection{Analysis on MeKB-Rec's Effectiveness}
The proposed MeKB-Rec method uses the personal knowledge graph, MeKB, as a medium for expressing cross-domain interests. As the method strikes CDR performance to a new level, we analyze why it is effective from two standpoints:


1) Domain-agnostic interest representation of MeKB. MeKB uses global entities to represent user interests. Entities are domain-agnostic elements that are semantically stable, therefore can naturally serve as units of interests that are valid across domains, say from Movies to Music, or from Articles to Videos.

2) Few-shot knowledge transfer of PLMs. Traditional CDR methods suffer from limited training examples of overlapping users. As shown in \autoref{appendix:freq_bucket}, they fail to generalize to users with no or few activities in the target domain. MeKB-Rec on the other hand, can benefit world knowledge learned from PLMs to understand and transfer the semantics to better serve recommendations in the target domain. The EmbedMe training procedure is essentially a few-shot fine-tuning step of the original PLM to fit the target recommendation task. As PLMs are good few-shot learners, this approach is especially effective. In \autoref{appendix:plm_ablation} we show that by removing the pretrained parameters, the performance drops by 25\%--40\%.

\subsection{Online Experiments}

\begin{figure}[t]
  \centering
  \includegraphics[width=1.0\linewidth]{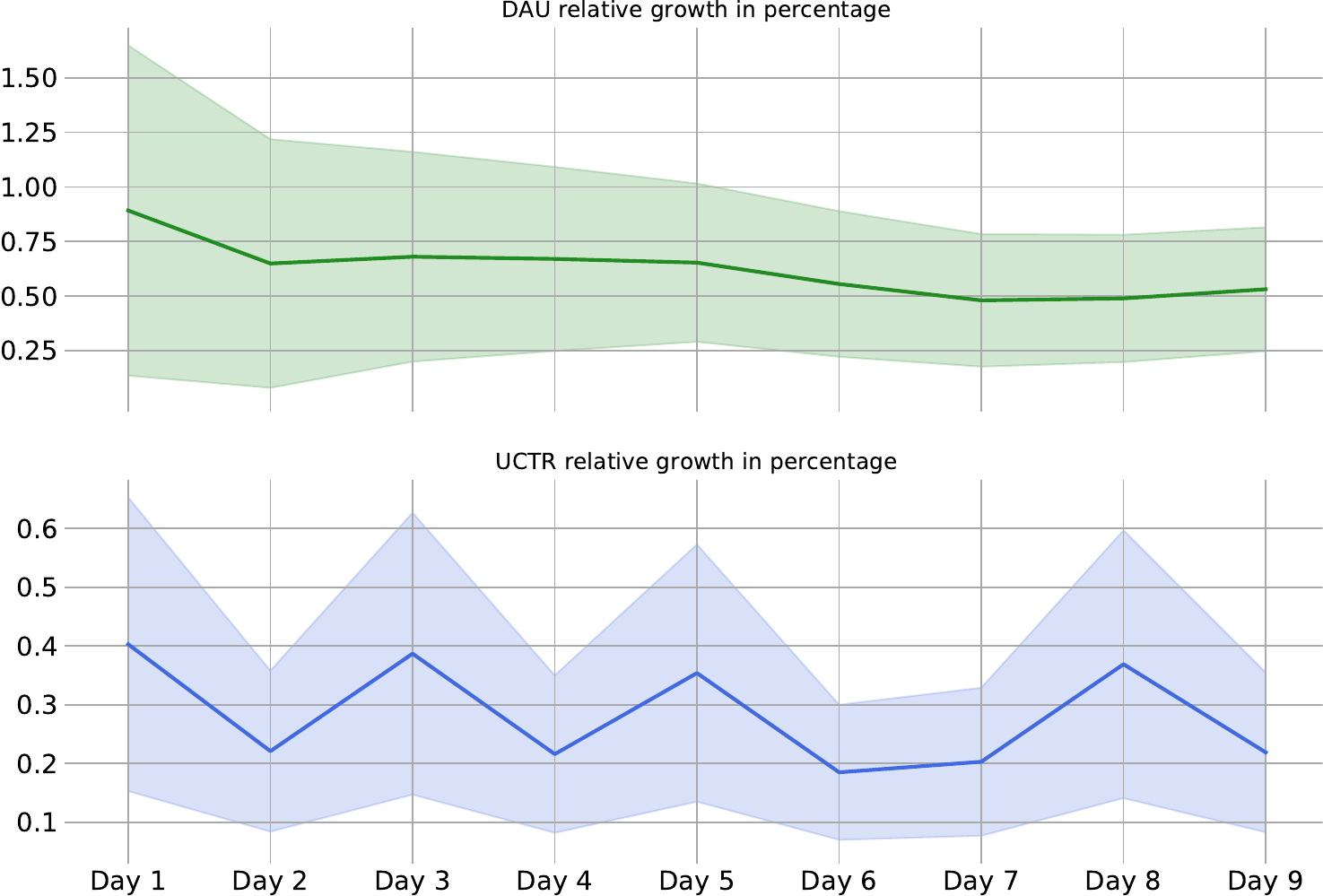}
  \caption{DAU and UCTR relative percentage change of an online A/B experiment using MeKB-Rec for WeiXin Live Recommendation.
  }
  \vspace{-0.2cm}
  \label{Fig-online-exp}
\end{figure}

We conduct A/B experiments in the WeiXin online recommendation system serving hundreds of millions of users. We report results in a major Live-stream recommendation scenario in the platform, where we deploy MeKB-Rec to leverage user behaviors in the short video domain to improve the recommendation for live streams, especially for cold-start users.

The control group is the current recommendation system in production. The experiment group is the same system with MeKB-Rec as an additional retrieval step for recommendation candidates.
Experiments are run for 9 days, during which the item tower of MeKB-Rec is continuously trained on an hourly basis, to fit recently-popular live broadcasters. 
We focus on two metrics: (1) DAU of non-active users in Live, and (2) Overall UCTR for all users, which is the unique click-through rate at the user level.

The experimental result is shown in \autoref{Fig-online-exp}. The X-axis is the day of the experiment, and Y-axis is the relative percentage improvement of the experiment group compared to the control. 
The metric shown in the DAU chart is accumulative while the UCTR chart shows a daily metric. The error bar stands for the 95\% confidence interval.
Overall, the experiment group with MeKB-Rec achieves a significant gain of DAU $+0.531\%$ ($p<0.01$) for non-active users, suggesting strong improvements over cold-start (zero-shot and few-shot) users. This pool of users accounts for a significant proportion and is critical for the growth of the recommendation scenario. The overall UCTR of all users is improved by $+0.240\%$ ($p<0.01$), indicating that users are generally more in favor of the recommended content compared to the baseline.

As online experiments show significant gains, MeKB-Rec has been deployed to the production system for 100\% of non-active users in this scenario.



%% file: 6-conclusion.tex
In this paper, we propose a novel knowledge-graph-based method, MeKB-Rec, that significantly improves Cross-Domain Recommendation (CDR) performance and reduces the requirement for overlapped users' information.
MeKB-Rec is a two-phase method, firstly constructing MeKB, a user's personal knowledge graph, 
and secondly mapping MeKB entities into a semantic embedding space for recommendations in the target domain, namely the EmbedMe algorithm.
MeKB-Rec achieves state-of-the-art recommendation results on multiple CDR scenarios in Amazon and WeiXin. 
The method also achieved significant gains in online experiments in WeiXin Live recommendation, and is deployed in the product serving hundreds of millions of users.

Specifically, MeKB-Rec has achieved benefits in the following aspects:
(1) obtained substantial advantages on large-scale CDR datasets: 38\%--67\% relative improvement of HR@10 on Amazon and 112\%--114\% gain of HR@10 on WeiXin datasets;
(2) utilized the information contained in the global KG, to supplement the feature representation for user interests that carry across domains;
(3) used PTMs to efficiently learn on insufficient training examples.
We conclude that injecting global semantics via KG and PTM largely eases learning when the data is sparse. This assumption is validated by MeKB-Rec's noticeable wins on zero-shot and few-shot user groups (respectively 105\% and 66\% HR@10 gains on Amazon) insufficient for traditional methods to learn a reliable user representation.

Regarding limitations, we note that the quality of the construction of MeKB is affected by the accuracy of the entity linking method. Specifically, the ambiguity of entity names or the insufficient coverage of entities will cause the representation deviation of MeKB. Future work may exploit more comprehensive entity linking models to improve MeKB's expressiveness and
MeKB-Rec's overall results.


The work of this paper is an important attempt at knowledge-graph-based methods in Cross-Domain Recommendations. Future work may focus on: (1) combining knowledge graphs with Large Language Models to further improve recommendations; (2) leveraging more sophisticated content understanding methods to improve the construction of MeKB; (3) better representation learning mechanisms to improve Embedme; and (4) exploring more usage for personal knowledge graphs such as MeKB.

%% file: 10-appendix1.tex
In this section, we show more details about datasets in the experiments.

\textbf{Amazon Datasets.} The Amazon dataset is a widely used public recommendation dataset. We selected three popular domains in Amazon: (1) Books, (2) Movies and TV, and (3) CDs and Vinyl, abbreviated as \textbf{Books, Movies and Music}. 
In the CDR problem, we use these three domains of data to construct three cross-domain scenarios. 
Following \citet{cao2022disencdr, liu2020cross}, two domains are combined into one CDR scenario.
We combine the domains into three CDR scenarios: \textit{Amazon Books $\rightarrow$ Movies}, \textit{Amazon Books $\rightarrow$ Music}, and \textit{Amazon Movies $\rightarrow$ Music}. 
The degree of the semantic gap between domains in these scenarios is different, which can verify the robustness of the methods tested.
Positive interactions are defined as a review score of 4 or above.
Following prior work \cite{cao2022disencdr}, we filter out users and items based on the frequency of their interactions. Specifically, we filter out users with less than five positive interactions and filter out items with <50 positive interactions as the focus of this work is not on cold-start items.
We randomly select 20\% of the overlapped users and place all their records into the test set, to verify our effectiveness on cold-start users.

\begin{table*}[t]
\centering
\caption{Statistics of the CDR datasets.}
\label{Tab-statis-01}
\begin{tabular}{cccccc}
\hline
\multicolumn{2}{c}{\textbf{Dataset}}   & Users   & \begin{tabular}[c]{@{}c@{}}Overlap\\ Ratio \end{tabular} & Items  & Records   \\ \hline
\multirow{2}{*}{\textbf{Amazon}} & Books & 163,189   & 4.77 $\%$   & 34,820  & 2,431,128   \\ 
                                 & Movies & 18,386   & 42.33 $\%$   & 24,621  & 707,037   \\ \hline
\multirow{2}{*}{\textbf{Amazon}} & Books  & 163,189   & 4.49 $\%$   & 34,820  & 2,431,128   \\
                                 & Music & 61,229   & 11.97 $\%$   & 24,621  & 707,037   \\ \hline
\multirow{2}{*}{\textbf{Amazon}} & Movies & 18,386   & 42.26 $\%$   & 24,621  & 707,037   \\
                                 & Music & 61,229   & 12.69 $\%$   & 55,219  & 826,490   \\ \hline
\multirow{2}{*}{\textbf{WeiXin}} & Articles   & 2,270,800 & 52.19$\%$   & 21,432 & 13,197,438   \\
                                 & Videos & 1,260,491 & 94.02$\%$    & 20,617 & 14,230,730   \\ \hline
\multirow{2}{*}{\textbf{WeiXin}} & Videos & 1,308,548 & 55.56$\%$   & 20,617 & 10,275,617   \\
                                 & Live  & 818,773 & 88.79$\%$    & 15,030 & 7,262,557   \\ \hline

\end{tabular}
\end{table*}

\textbf{WeiXin Datasets.} We build large-scale CDR datasets extracted from WeiXin recommendation scenarios.
The dataset comes from real user-item interaction on the WeiXin platform, involving three recommendation domains: Articles, Channels (short videos), and Live (live broadcasts).
We set up two cross-domain recommendation scenarios that are \textit{WeiXin Articles $\rightarrow$ Videos} and \textit{WeiXin Videos $\rightarrow$ Live},
and construct CDR datasets with millions of records. 
To construct this dataset, we consider reading an article, fully watching a video, and watching a live stream for $\ge 30$ seconds as positive behaviors. We randomly sample all positives within two months for training and the week after for testing. 
Similarly, we filter out items with <50 positives. 
When constructing the data set, we anonymized the user information, and only kept the user's anonymous id and sampling behavior samples required for the experiment.
Compared with the Amazon dataset, the WeiXin dataset covers a much larger scale of users and items.
The detailed parameters of the Amazon and WeiXin datasets used in the experiments are shown in \autoref{Tab-statis-01}. 
The Overlap Ratio is the fraction of overlapped users of two domains over the number of users in the current domain, i.e. $OR(X, Y) = |U_{X} \cap U_{Y}| / |U_{X}|$.


\textbf{Dataset-specific KG augmentation.}
Although Wikidata is a general KG with rich world information, complementing it with domain-specific knowledge can be helpful for the recommendation.
We further incorporate dataset-specific entities into the KG to be considered in building MeKB, to increase item coverage and enrich user representations.
For the Amazon dataset, we include three types of entities available in the dataset: brands, categories, and products, each represented by its textual name. They are considered as additional entities with a lower weight (set to 0.5, 0.3 and 0.1 respectively, where original Wikidata entities are assigned a weight of 1.0)
Entity Linking on these entities is trivial, as each item in the dataset already have these entities denoted.
For the WeiXin dataset, we include an additional keyword dictionary as domain-specific entities, such as hashtags in videos. Custom entities are associated with items via keyword matching.

Following prior work \cite{cao2022disencdr, kang2019semi-sscdr}, examples in both source and target domains can be used in CDR evaluation. Therefore we combine interactions in both domains with equal weights to construct a user's MeKB, though in practice one can choose to only use the source domain to construct a pure out-of-domain MeKB for specific tasks. 
We only use examples in the training set to construct MeKB to make sure there is no data leakage on validation or test examples.

%% file: 11-appendix2.tex
\subsection{Impact of Pretrained Language Models}
\label{appendix:plm_ablation}

\begin{table*}[t]
\centering
\caption{Impact of Pretrained Language Models in different CDR scenarios in the Amazon dataset.}
\vspace{-0.2cm}  
\label{Tab-amazon-plm}
\begin{tabular}{lcccccc}
\toprule
\multirow{2}{*}{Methods} & \multicolumn{2}{c}{\begin{tabular}[c]{@{}c@{}} \textbf{Scenario 1}\\
    Books $\rightarrow$ Movies  
\end{tabular} }  
& \multicolumn{2}{c}{\begin{tabular}[c]{@{}c@{}} \textbf{Scenario 2}\\
    Books $\rightarrow$ Music  
\end{tabular} }
& \multicolumn{2}{c}{\begin{tabular}[c]{@{}c@{}} \textbf{Scenario 3}\\
    Movies $\rightarrow$ Music  
\end{tabular}} \\
\cmidrule(r){2-3} \cmidrule(r){4-5} \cmidrule(r){6-7}   & HR@10  & NDCG@10  & HR@10  & NDCG@10  & HR@10  & NDCG@10 \\
\midrule
w/o PLM  & 0.1539 & 0.0818   & 0.3667 & 0.2283  & 0.3430 & 0.2118  \\
MeKB-Rec & 0.2471 & 0.1370   & 0.4884 & 0.3298  & 0.5035 & 0.3444 \\
\midrule
Improved  & \textbf{+60.56\%} & \textbf{+67.48\%}  & \textbf{+33.18\%}  & \textbf{+44.45\%}   & \textbf{+46.79\%}  & \textbf{+62.61\%}   \\
\bottomrule
\end{tabular}
\vspace{-0.15cm}
\end{table*}

\begin{table*}[t]
\centering
\caption{Frequency-binned analysis on user activity. We split users in the Movies test set in the Amazon Books $\rightarrow$ Movies scenario, based on their frequency seen in training data, and report HR@10 numbers in each bucket. This analysis measures recommendation performance for users with limited or no behaviors for different approaches.}
\label{Tab-user-frequency}
\begin{tabular}{lcccccccc}

\toprule
\begin{tabular}[c]{@{}c@{}}Action Frequency\\ (\# of Users) \end{tabular} & \multicolumn{2}{c}{ \begin{tabular}[c]{@{}c@{}}Zero-shot Users: $[0]$\\ (1,562) \end{tabular} }        & \multicolumn{2}{c}{\begin{tabular}[c]{@{}c@{}}Few-shot Users: $[1, 10)$\\  (4,201) \end{tabular}} &   \multicolumn{2}{c}{\begin{tabular}[c]{@{}c@{}}Multi-shot Users: $[10, +\infty)$\\ (12,623) \end{tabular}}       & \multicolumn{2}{c}{ All Users }     \\
 \cmidrule(r){2-3} \cmidrule(r){4-5}  \cmidrule(r){6-7} \cmidrule(r){8-9}       & HR@10  & NDCG@10 & HR@10 & NDCG@10 & HR@10 & NDCG@10 & HR@10 & NDCG@10  \\
\midrule
YouTubeDNN & 0.0041 & 0.0003 & 0.1225 & 0.0784 & \underline{0.2512} & \underline{0.1402} & 0.1271 & 0.0762 \\
CMF               & 0.0642    & 0.0315  & 0.1433   & 0.0696    & 0.1912    & 0.0907 & 0.1726 & 0.0995  \\
CoNet & 0.0423 & 0.0196 & 0.1474 & 0.0610 & 0.1735 & 0.0861 & 0.1364 & 0.0835 \\
EMCDR             & {0.0782}    & \underline{0.0589} & \underline{0.1545}   & \underline{0.0815}    & 0.2008    & 0.0945 & 0.1755 & \underline{0.1103}  \\
DCDCSR & 0.0526 & 0.0345 & 0.1360 & 0.0638 & 0.1807 & 0.0811 & 0.1450 & 0.0826 \\
DDTCDR             & 0.0573    & 0.0323  & 0.1420   & 0.0704    & 0.1971    & 0.0911 & 0.1717 & 0.0805   \\
SSCDR             & \underline{0.0841}    & 0.0521  & 0.1478   & 0.0775    & {0.2072}    & {0.1053}  & \underline{0.1796} & 0.0919 \\
\midrule
\textbf{MeKB-Rec} & \textbf{0.1724} & \textbf{0.0936}  & \textbf{0.2563}   & \textbf{0.1655}    & \textbf{0.2554}    & \textbf{0.1411}  & \textbf{0.2471} & \textbf{0.1370} \\
Improved  & \textbf{+104.99\%} & \textbf{+58.91\%}  & \textbf{+65.89\%}   & \textbf{+103.07\%}    & \textbf{+1.67\%}    & \textbf{+0.64\%}  & \textbf{+37.58\%}    & \textbf{+24.21\%} \\
\bottomrule
\end{tabular}  
\end{table*}

As mentioned above, MeKB has a strong representation capability of users' interests, but this capability also brings the challenge of feature mapping for recommendation. Our proposed MeKB-Rec alleviates the problems of insufficient training records with the help of a pre-trained language model. In essence, MeKB embodies a large amount of world knowledge contained in the global KG to build a complete expression of users' interests. 
The training records in recommendation scenarios cannot cover the semantic information of all entities in the global KG. 
The introduction of the pre-trained language model enables the semantic understanding of entities in the global KG by pre-training on massive external corpora.

This hypothesis is confirmed in an ablation study shown in \autoref{Tab-amazon-plm}. We observe a significant decline in recommendation performance in MeKB-Rec, after dropping PLM parameters obtained during pretraining.
The ``w/o PLM'' setting uses the same model architecture but replaces the PLM parameters with random initialization. 
This indicates that the world semantic knowledge obtained during pretraining is critical for modeling users' interest in our model architecture.
We suggest that future work may try to further pretrain the PLM on the corpus from the recommendation scenarios to improve the adaptation effect to the recommendation system.

\subsection{Frequency-binned Analysis on User Activity}
\label{appendix:freq_bucket}

To understand how well our method performs for users with a limited amount of training data, we split users in the Movies test set based on their frequency seen in training data in the target domain, and report HR@10 numbers in each bucket. 
As shown in \autoref{Tab-user-frequency}, MeKB-Rec achieves uniform improvement in all user groups compared to all baselines.
Particularly, MeKB-Rec shows a significant advancement for the zero-shot user group (appearing zero times in the training set) and few-shot users (seen less than ten times in training) compared to other methods, with \textbf{105\%} and \textbf{66\%} HR@10 gains respectively.
MeKB-Rec performs less well on high-activity users with more than 10 training examples, as the baseline methods usually have a dedicated ID embedding that fits these users better after seeing many examples.
As expected, MeKB-Rec is better for zero-shot and few-shot users which is a critical challenge in CDR that we aim to address.